\documentclass[sigconf]{acmart}

\usepackage{graphicx}
\usepackage{textcomp}
\usepackage{amsmath}
\usepackage{listings}

\usepackage{tikz}
\usetikzlibrary{tikzmark}
\usepackage{subcaption}
\usepackage{xcolor}
\usepackage{capt-of}

\usepackage{multirow}

\usepackage{url}
\usepackage{booktabs}

\lstdefinestyle{mystyle0}{
    escapeinside={(*@}{@*)},
}

\lstdefinestyle{mystyle}{
  language=C, 
  commentstyle=\color{green!60!black}, 
  literate=*{x_1}{{\texttt{x$_1$}}}2 {x_2}{{\texttt{x$_2$}}}2 {x_3}{{\texttt{x$_3$}}}2,
}

\lstdefinestyle{mystyle3}{
    language=C,
    basicstyle=\ttfamily,
    keywordstyle=\color{green!60!black},
    commentstyle=\color{green!60!black},
    breaklines=true,
    showstringspaces=false,
    escapeinside={(*@}{@*)},
}

\AtBeginDocument{%
  \providecommand\BibTeX{{%
    \normalfont B\kern-0.5em{\scshape i\kern-0.25em b}\kern-0.8em\TeX}}}

\setcopyright{acmcopyright}
\copyrightyear{2018}
\acmYear{2018}
\acmDOI{XXXXXXX.XXXXXXX}

\acmConference[Conference acronym 'XX]{Make sure to enter the correct
  conference title from your rights confirmation emai}{June 03--05,
  2018}{Woodstock, NY}
\acmPrice{15.00}
\acmISBN{978-1-4503-XXXX-X/18/06}

\begin{document}

\title{Multi-structure Objects Points-to Analysis}

\author{Xun An}


\orcid{1234-5678-9012}
\email{anxun@iie.ac.cn}




\renewcommand{\shortauthors}{Xun An, et al.}

\begin{abstract}
An important dimension of pointer analysis is field-sensitivity, 
which has been proven to effectively enhance the accuracy of pointer analysis results. 
A crucial area of research within field-sensitivity is structure-sensitivity. 
Structure-sensitivity has been shown to further enhance the precision of pointer analysis. 
However, existing structure-sensitive methods cannot handle cases where an object possesses multiple structures, 
even though it's common for objects to have multiple structures throughout their lifecycles. 
Our observation confirmed that it's common for objects to have multiple structures throughout their lifecycle.

This paper introduces MTO-SS, a flow-sensitive pointer analysis method for objects with multiple structures. 
MTO-SS ensures that the structural information of an object is always accurate and complete at different positions in the program by introducing structure-flow-sensitivity. 
MTO-SS always performs weak updates on the structure of an object. 
This means that once an object has a structure, that structure will accompany the object throughout its entire lifecycle. 
We evaluated our multi-structure object pointer analysis method using the 12 largest programs from GNU Coreutils. 
We then compared the experimental results with the fully sparse flow-sensitive method, SPARSE, and another method, TYPECLONE, which only allows an object to have a single type. 
Our experimental results confirm that MTO-SS is more precise than both SPARSE and TYPECLONE. 
On average, it can answer over 22\% more alias queries with no-alias results compared to SPARSE, and over 3\% more compared to TYPECLONE. 
Additionally, the time overhead introduced by our method is very low.
\end{abstract}






\maketitle

\section{Introduction}

Pointer analysis statically determines the objects that a pointer may point to at runtime. 
Pointer analysis techniques serve as foundational research tools for various other analyses, 
such as vulnerability detection, compiler optimization, program comprehension, etc. \cite{diwan1998type, le2005using, krishnamurthi2016boomerang, livshits2003tracking}. 
Field-sensitivity can significantly enhance the accuracy of pointer analysis results, 
and the improvement in accuracy brought by field-sensitivity has been demonstrated in Java language pointer analysis \cite{berndl2003points, rountev2001points, milanova2002parameterized, bravenboer2009strictly}. 
Although field-sensitivity has shown remarkable accuracy enhancement in Java pointer analysis, there is limited research on applying field-sensitivity to C/C++ pointer analysis \cite{hardekopf2007ant, pereira2009wave, hardekopf2007exploiting, blackshear2011flow, zhang2014efficient, lei2019fast}. 
This is because applying field-sensitivity in C/C++ pointer analysis is more challenging than in Java. 
As previous studies have pointed out \cite{pearce2007efficient, balatsouras2016structure, lei2019fast}, the key challenge is that C/C++ allows retrieving the address of a field and directly accessing memory using that address. 
However, Java does not permit obtaining the address of a field, and in the Java language, accessing a field requires an additional type identifier to specify its type. 
In C/C++, memory access only requires an address without any additional identifiers. 
Therefore, when conducting field-sensitive analysis for C/C++, the analysis method itself must determine the field of the object being accessed.

For C/C++ field-sensitive pointer analysis, a representative method is the field-index-based approach proposed by Pearce et al. \cite{pearce2007efficient,balatsouras2016structure, lei2019fast}. 
This method differentiates fields of an object by assigning a unique index to each field to achieve field-sensitivity. 
The method extends the constraint graph used in Andersen's pointer analysis \cite{andersen1994program}. 
In the original graph, nodes denote variables, while edges depict the relationships between these variables. 
Pearce et al. proposed extending the constraint graph by adding positive weights to its edges, where these weights represent the indices of the object's fields. 
Subsequently, this field-sensitive analysis technique determines the fields pointed to by pointers on the expanded constraint graph. 
Later researchers proposed improvements to this method to solve the constraint graph faster \cite{lei2019fast}. 
However, since these techniques do not take into account the structural information of objects \cite{balatsouras2016structure,barbar2020flow}, two issues arise: 
1. These methods introduce a new cycle on the constraint graph, called the positive weight cycle (PWC), which is notably difficult to address. 
2. The absence of structural information results in numerous spurious points-to inferences.

cclyzer-ss \cite{balatsouras2016structure} utilizes object structural information to enhance the accuracy of pointer analysis to address the issues present in Pearce et al.'s methods. 
cclyzer-ss integrates the structural information of objects, utilizing it to filter out spurious points-to relations. 
However, cclyzer-ss can only handle cases where an object has a single type throughout its entire lifecycle, 
and it results in erroneous points-to relations when processing objects that possess multiple types throughout their lifecycle. 
As pointed out in \cite{avots2005improving}, it is common for objects to have multiple types throughout their lifecycle, and our experiments further confirmed this. 
TYPECLONE \cite{barbar2020flow} combines structure-sensitivity with flow-sensitivity, further improving the accuracy of pointer analysis. 
TYPECLONE shares the same limitation with cclyzer-ss that it cannot handle the case where an object has multiple types. 
Moreover, both cclyzer-ss and TYPECLONE neglect the dynamic type conversion function $dynamic\_cast$ in C++. 
Both methods merely treat the dynamic type conversions as simple assignment statements, leading to incorrect points-to result.

To overcome the limitations of current designs that inability to correctly analyze cases where objects have multiple structures, 
and that are unable to handle the dynamic type conversion function $dynamic\_cast$, we present MTO-SS in this paper. 
MTO-SS is a structure-flow-sensitive pointer analysis method that can quickly and accurately analyze the cases where an object has multiple structures. 
MTO-SS introduces the flow sensitivity of structures for objects, which ensures that the structure information of an object at diverse program locations remain comprehensive and exact. 
Furthermore, MTO-SS isn't limited to just static type conversions. 
It also capably processes dynamic type conversions caused by C++'s $dynamic\_cast$ function. 
MTO-SS utilizes the structural information of objects to increase the accuracy of points-to results. 
We implemented MTO-SS based on SVF-2.5 and our source code is available at \url{https://github.com/Frankenstein-bit/Structure-Flow-Sensitive}. 
We evaluated our approach using the 12 largest programs in GUN Coreutils. 
The experimental results indicate that, compared to the fully sparse flow-sensitive analysis SPARSE \cite{hardekopf2011flow}, 
the accuracy of our method has improved by an average of 22.94\%. 
In comparison to the TYPECLONE method \cite{barbar2020flow}, which assumes objects have only one type, the accuracy of our method has also increased by an average of 3.23\%. 
Our experimental results show that our method is very effective.

Our contributions are as follows: 

1. We have proposed a pointer analysis method, MTO-SS, which can analyze objects with multiple structures. 
MTO-SS introduces structure-flow-sensitivity of object and utilizes structural information to enhance the accuracy of the pointer analysis results. 

2. MTO-SS is not only capable of analyzing static type conversions but can also deal with dynamic type conversion caused by C++ $dynamic\_cast$ function. 

3. We have implemented MTO-SS and compared it with SPARSE, as well as the TYPECLONE method which assumes objects have only one type. 
Experimental results show that the accuracy of MTO-SS is, on average, 22.94\% higher than SPARSE, and 3.23\% higher than that of TYPECLONE.

\section{Motivation Examples}

\begin{figure}[htbp]
\centering

\begin{lstlisting}[numbers=left, numberstyle=\normalfont, numbersep=-8pt, commentstyle=\color{green!60!black}, language=C++]
    typedef struct{
        char s1;
        char s2[4];
    } T1;
    typedef struct{
        char st[6];
    } T2;
    void main() {
      T1* p = (T1*)malloc(5); // o                      
      p->s1 = '1';
      memcpy(p->s2,"asa",4);
      T2* q = (T2*)p;
      char* str =  q->st;
    }
\end{lstlisting}

\centering
\caption{A multi-structure object example.}
\label{fig:motivationA}

\end{figure}

We highlight the shortcomings of current structure-sensitive solutions using two examples depicted in Figure \ref{fig:motivationA} and Figure \ref{fig:motivationB}. 
In Figure \ref{fig:motivationA}, an object $o$ is allocated in line 9 using the $malloc$ function and is cast to type $T1$. Subsequently, in line 12, it is recast to type $T2$. 
Existing solutions cclyzer-ss and TYPECLONE instantiate a distinct object for each type encountered. 
For instance, in this case, cclyzer-ss and TYPECLONE would instantiate an object $o'$ for type $T1$ and associate $T1$ with $o'$. 
Similarly, they would establish a new object $o''$ for type $T2$ and associate $T2$ with $o''$. 
Furthermore, they would continue propagation using the newly created objects $o'$ and $o''$ in place of the original object $o$. 
These existing methods assume that objects $o'$ and $o''$ are two unrelated objects. 
Thus, when performing alias queries, the existing methods infer that the pointers $p$ and $q$ are not aliases. 
Similarly, they infer that the fields $q$->$s1$ and $q$->$st$ are not aliases, leading to erroneous results.

In Figure \ref{fig:motivationB}, four classes are declared: $B$ is the base class, and both $L$ and $R$ inherit from class $B$. 
$D$ is derived from both $L$ and $R$. 
Since classes B, L, and R each include virtual functions, each of these classes contains only one virtual table pointer in memory. 
Because class D inherits from both classes L and R, it has two virtual table pointers in memory: the first one points to class L's virtual table, and the second points to class R's virtual table. 
An object $o1$ of class $D$ is created in line 6, only containing two virtual table pointers within it. 
In the code on lines 7 and 8, the pointers $left$ and $baseL$ both point to the start of $o1$, where the first pointer is located. 
In the code on line 10, through the $dynamic\_cast$ conversion, the pointer $ri$ points to the location of the second pointer of the object $o1$. 
Thus the $f()$ function of class $R$ is called at line 11. 
However, the existing approach neglects this case and simply treats the $dynamic\_cast$ type conversion as a copy instruction. 
As a result, existing methods assume that the pointer $ri$ points to the first virtual table pointer of the object $o1$, leading to incorrect analysis results. 

\begin{figure}[htbp]
\centering
\begin{lstlisting}[numbers=left, numberstyle=\normalfont, numbersep=-8pt, commentstyle=\color{green!60!black}, language=C++]
    struct B {virtual void f(){}};
    struct L:B {virtual void f(){}};
    struct R:B {virtual void f(){}};
    struct D : L, R { };
    void main(){
      D* der = new D(); //o1
      L* left = der;
      B* baseL = left;
      baseL->f();   //call L:f()
      R* ri=dynamic_cast<R*>(baseL);
      ri->f();     //call R:f()
    }

\end{lstlisting}
\centering
\caption{A $dynamic\_cast$ type conversion example.}
\label{fig:motivationB}
\end{figure}

  \begin{table*}[htbp]
    \caption{Domains and LLVM-like instructions used by our pointer analysis}
    \label{tab:inst}
    \begin{tabular}{lll|ll}
    \hline
    \multicolumn{3}{c|}{Analysis domains}                                                                                 & \multicolumn{2}{c}{LLVM-like instruction set}                      \\ \hline
    $\ell$            & $\in$    $\mathcal{L}$                                                  & instruction labels      & $p = alloca \ t,nbytes$                   & STK/GLOBAL      \\
    $t$               & $\in$    $\mathcal{T}$                                                  & program structures                  & $p = malloc \ nbytes$                     & HEAP       \\
    $off$              & $\in$    $\mathcal{C}$                                                  & field offsets      & $p = q$                                   & COPY                  \\
    $fld$           & $\in$    $\mathcal{F}$                                                    & fields  & $p = (t)q$                                & CAST                  \\
    $c,k$             & $\in$    $\mathcal{C}$                                                  & constants                   & $p = *q$                                  & LOAD     \\
    $s$               & $\in$    $\mathcal{S}$                                                  & virtual registers               & $*p = q$                                  & STORE       \\
    $g$               & $\in$    $\mathcal{G}$                                                  & global variables & $p = \phi(q,h)$                           & PHI           \\
    $m$               & $\in$    $\mathcal{M}$ $\subseteq$  $\mathcal{G}$                       & program functions        & $p = \&q[idx]$                            & ARRAY \\
    $p,q,r,x,a$       & $\in$    $\mathcal{P}$ $=$ $\mathcal{S}$ $\cup$  $\mathcal{G}$          & top-level variables              & $p = \&q \rightarrow fld$                 & FIELD       \\
    $o$               & $\in$    $\mathcal{O}$                                                  & abstract objects     & $constructor(t \ o,...)$                   & CONSTRUCTOR-C++ \\
    $\widehat{o_{off}}$ & $\in$    $\mathcal{O \times F}$                                         & abstract fields of objects         & $p = dynamic\_cast\langle t*\rangle (q)$  &DYNAMIC\_CAST-C++         \\
    $c,d$             & $\in$    $\mathcal{A}$ $=$ $\mathcal{O}$ $\cup$  $\mathcal{O\times F}$ & address-taken variables         & $p =q (...,a,...)$ & CALL          \\
    $v$               & $\in$    $\mathcal{V}$ $=$ $\mathcal{P}$ $\cup$  $\mathcal{A}$          & program variables & $ret \ p$                                 & RET       \\ \hline
    \end{tabular}
    \end{table*}

\section{Program Representation}

\subsection{Program Representation}

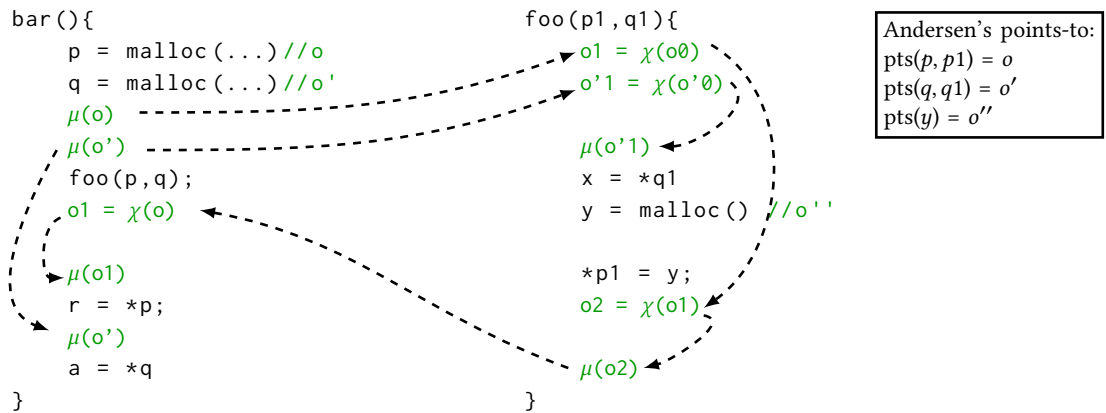
\begin{figure*}[htbp]
\centering
\begin{minipage}{0.4\textwidth}
\centering
\begin{lstlisting}[style=mystyle3]
      bar(){
          p = malloc(...)//o
          q = malloc(...)//o'
          (*@\color{green!60!black}$\mu$(o)@*)
          (*@\color{green!60!black}$\mu$(o')@*)
          foo(p,q);
          (*@\color{green!60!black}o1 = $\chi$(o)@*)

          (*@\color{green!60!black}$\mu$(o1)@*)
          r = *p;    
          (*@\color{green!60!black}$\mu$(o')@*)
          a = *q
      }
\end{lstlisting}
\end{minipage}
\begin{minipage}{0.4\textwidth}
\centering
\begin{lstlisting}[style=mystyle3]
    foo(p1,q1){
        (*@\color{green!60!black}o1 = $\chi$(o0)@*)
        (*@\color{green!60!black}o'1 = $\chi$(o'0)@*)

        (*@\color{green!60!black}$\mu$(o'1)@*) 
        x = *q1
        y = malloc() //o''

        *p1 = y;
        (*@\color{green!60!black}o2 = $\chi$(o1)@*)

        (*@\color{green!60!black}$\mu$(o2)@*)
    }
\end{lstlisting}
\end{minipage}

\begin{tikzpicture}[remember picture, overlay, line width=1pt, >=latex]
\draw[dashed, black, ->] (-4.3,4.2) to[out=0, in=-160] (1.5,5.0);
\draw[dashed, black, ->] (-4.2,3.7) to[out=0, in=-160] (1.5,4.5);
\draw[dashed, black, ->] (3.56,4.6) to[out=-40, in=0] (2.6,3.75);
\draw[dashed, black, ->] (3.3,5.1) to[out=-30, in=35] (3.2,1.6);
\draw[dashed, black, ->] (3.2,1.5) to[out=-10, in=15] (2.4,0.8);
\draw[dashed, black, ->] (1.4,0.8) to[out=160, in=-10] (-3.5,2.9);
\draw[dashed, black, ->] (-5.35,2.8) to[out=-160, in=-180] (-5.3,2.0);
\draw[dashed, black, ->] (-5.4,3.7) to[out=-120, in=160] (-5.5,1.3);
\node[draw, text width=2.8cm]  at (7.0,4.7) {Andersen's points-to: \\ pts($p,p1$) = {$o$} \\ pts($q,q1$) = {$o'$} \\ pts($y$) = {$o''$} };

\end{tikzpicture}

\caption{An inter-procedural value-flow example.}
\label{fig:InterObjectSSA}
\end{figure*}


Our structure-flow-sensitive pointer analysis is based on the LLVM IR of the program. 
LLVM IR is a strongly typed language that disallows implicit type conversions. 
This means that when a program performs memory access, the type of the pointer needs to match the type of the accessed object. 
LLVM defines a rich set of RISC instructions, but the instructions and domains related to our pointer analysis are already listed in Table \ref{tab:inst}. 
The variables $\mathcal{V}$ in the program can be classified into two types: address-taken variables $\mathcal{A}$, which are accessed through pointers, 
and top-level variables $\mathcal{P}$, which are never accessed through pointers. 
The address-taken variable $\mathcal{A}$ includes abstract objects $\mathcal{O}$ and the fields of abstract objects $\mathcal{O \times F}$. 
Top-level variables can be accessed directly, while address-taken variables can only be accessed through top-level variables, using load and store instructions.

LLVM defines a rich instruction set, but only a small subset of instructions are relevant to our structure-flow-sensitive pointer analysis. 
Table \ref{tab:inst} presents the instruction set that is related to our pointer analysis. 
The program is represented by 13 instructions, where 11 instructions are statements within the function body, and two instructions are entry and exit statements of the function. 
The STK/GLOBAL instruction allocates stack or global objects, with parameters $nbytes$ and $t$ denoting the object's size and type respectively. 
The HEAP instruction, used for heap object allocation, has a single parameter specifying the object's size. 
COPY is an assignment statement. 
PHI is the confluence point of branch instructions, with different branch statements choosing different values. 
CAST is used to interpret an object according to a specific type to satisfy LLVM's type checking without altering any content in memory. 
LOAD and STORE are read and write operations for address-taken objects. 
FIELD obtains the address of a field of an object. 
ARRAY returns the address of an element of an array. 
CONSTRUCTOR-C++ is a constructor function in C++ used for initializing the object and its RTTI (Run-Time Type Information) \cite{stroustrup1992run} information. 
DYNAMIC\_CAST-C++ is used in C++ to convert a base class object into an inherited class object or to perform object type conversion between sibling classes, and checks the legality of type conversion. 
The function call instruction CALL and function return instruction RET treat the passing of function parameters and return values as COPY operations.

The differences between C++ and C mainly involve two points: 
(1) The allocation of C++ objects, such as $Base *p = new \ \ Base()$, is translated into two LLVM instructions. 
One is a memory allocation instruction for allocating memory space for the object, and the other is a constructor call instruction to initialize the object. 
(2) $dynamic\_cast$ is a type conversion instruction specific to C++ that checks the legality of the object's type conversion. 
$dynamic\_cast$ checks whether there is an inheritance relationship between the converted type and initial type of the object. 
Moreover, $dynamic\_cast$ does not merely interpret the object as another type; it adjusts the location the pointer points to, based on the relationship between the object's current type and the type it is being converted into.

For pointer arithmetic, for example, $p = q + j$, if the pointer $q$ points to the object $o$, we conservatively assume that the pointer $p$ also points to the object $o$. 
This is based on the assumption that pointer arithmetic will not cross the boundary of the object.

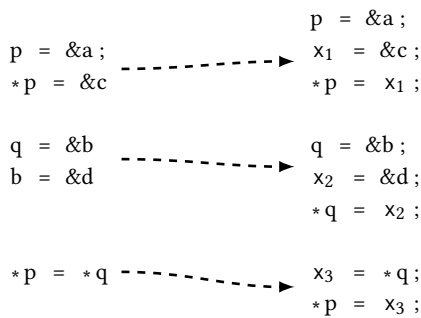
\begin{figure}[htbp]
\centering
\begin{minipage}{0.22\textwidth}
\begin{lstlisting}[style=mystyle]
    p = &a;
    *p = &c

    q = &b
    b = &d 


    *p = *q
\end{lstlisting}
\end{minipage}
\begin{minipage}{0.22\textwidth}
\begin{lstlisting}[style=mystyle]
    p = &a;
    x_1 = &c;
    *p = x_1;

    q = &b;
    x_2 = &d;
    *q = x_2;

    x_3 = *q;
    *p = x_3;
\end{lstlisting}
\end{minipage}

\begin{tikzpicture}[remember picture, overlay, line width=1pt, >=latex]
\draw[dashed, black, ->] (-1.7,3.5) to[out=0, in=180] (0.6,3.6);

\draw[dashed, black, ->] (-1.7,2.3) to[out=0, in=180] (0.6,2.2);

\draw[dashed, black, ->] (-1.7,0.8) to[out=0, in=180] (0.6,0.6);


\end{tikzpicture}

    \caption{C code fragment and Corresponding LLVM IR.}
    \label{fig:SSA}
\end{figure}

LLVM IR is in partial SSA form. 
Top-level variables are converted to SSA form using the standard SSA construction algorithm \cite{bilardi2003algorithms,chow1996effective,aycock2000simple}. 
Address-taken variables are not transformed into SSA form because they can only be accessed through top-level variables in load and store instructions. 
Figure \ref{fig:SSA} presents a section of code and its corresponding partial SSA form. 
In the example, $*p=*q$ is decomposed into two statements, introducing the top-level variable $x_3$.

\subsection{Value-flow Representation for Flow-sensitive Analysis}
\label{sec:valueflow}

We represent the program in the form of value-flow graph \cite{hardekopf2011flow, oh2012design, shi2018pinpoint, sui2016demand}, and we perform structure-sensitive pointer analysis based on flow-sensitive pointer analysis. 
Unlike flow-insensitivity, which ignores the execution order of the program, flow-sensitivity considers the execution order of the program \cite{li2011boosting,hardekopf2011flow}. 
Traditional flow-sensitive methods maintain points-to relationships for all variables at each point in the program based on the Control Flow Graph (CFG) \cite{landi1992safe}. 
These methods propagate the points-to relationships of each point in the CFG in an iterative manner until a fixed point is reached \cite{choi1993efficient, emami1994context, kam1977monotone, wilson1995efficient}. 
As these methods cannot determine whether each variable is useful at each point, they can only conservatively propagate all points-to information for each variable at each point, 
leading to significant time and memory consumption.

To accelerate flow-sensitive analysis, modern methods generally employ a sparse approach for propagating points-to relationships. 
The value-flow graph (VFG) is a sparse def-use graph where each node represents a program statement and edges represent def-use relationships of variables. 
VFG directly connects variable definition statements to variable use statements, avoiding the blind propagation of unrelated points-to relationships. 
We represent the program in the form of the value-flow graph. 

The top-level variables of the program are directly in SSA form on LLVM IR, and their def-use relationships are directly available without additional pointer analysis. 
Edges formed by def-use relationships between top-level variables are called direct edges. 
For example, $\ell \xrightarrow{p} \ell'$ and $p \in \mathcal{P}$, indicates that variable $p$ is defined at program point $\ell$ and used at $\ell'$. 
Due to LLVM's partial SSA form, address-taken variables are not transformed into SSA form. 
These variables are defined through store instructions and used through load instructions. 
We take several steps to construct SSA for address-taken variables. 
(1) We use fast flow-insensitive Andersen's pointer analysis \cite{andersen1994program, hardekopf2007ant, pereira2009wave} to compute the points-to information of the program. 
(2) We implement SSA for address-taken variables by annotating load and store instructions. 
For load instructions like $p = *q$, where the pointer $q$ points to object $o$ and $o \in \mathcal{O}$, we annotate this load instruction with the function $\mu(o)$, 
indicating the use of variable $o$ in this statement. 
For store instructions like $*p = q$, where the pointer $p$ points to $o$ and $o \in \mathcal{O}$, we use $o =\chi(o)$, indicating that the address-taken variable $o$ is both defined and used in this statement. 
This means the store instruction is equivalent to first using the address-taken variable and then defining it. 
(3) We treat $\mu(o)$ as the use of $o$, and $o =\chi(o)$ as the definition and use of $o$, and then transform them into SSA based on the position of address-taken variables in the CFG. 
Finally, we obtain the def-use relationships of the address-taken variables. 
For instance, an indirect def-use relationship $\ell' \xrightarrow{o} \ell$ and $o \in \mathcal{O}$, indicates that the object $o$ is defined and used at $\ell'$ and used at $\ell$. 

We also consider the inter-procedural def-use relationships of address-taken variables \cite{hardekopf2011flow, sui2014detecting}. 
To handle the impact of object usage within functions, we add $\mu$ function before the call point to indicate that the object is used, and we add $\chi$ function at the start of the corresponding function to indicate that the function has received the object. 
For example, in Figure \ref{fig:InterObjectSSA} for the call to function $foo$, we add usage functions $\mu(o)$ and $\mu(o')$ for objects $o$ and $o'$ before the call point $foo(p,q)$. 
At the start of the function $foo$, we add definition functions $o1=\chi(o0)$ and $o'1=\chi(o'0)$ for these two address-taken variables. 
Similarly, to address the impact of object definition within a function, we add $\mu$ function at the return point to return the object, and after the call point we add $\chi$ function to receive the object returned from the return point. 
In Figure \ref{fig:InterObjectSSA}, at the return position of function $foo$, we add a usage function $\mu(o2)$ for object $o$, and after the call point $foo(p,q)$ in function $bar$, we add a definition function $o1=\chi(o)$ for object $o$. 
We perform Andersen's pointer analysis within the function to determine the use or definition of objects. 
Finally, we convert the address-taken variables between procedures into SSA form in a similar way to those within procedures. 
The dashed line in Figure \ref{fig:InterObjectSSA} represents the def-use chains of objects $o$ and $o'$.

\section{MTO-SS Approach}

In this section, we will first introduce our type model, then present our structure-flow-sensitive pointer analysis inference rules. 
Our multi-structure object pointer analysis inference rules have already been shown in Figure \ref{fig:inferenceFig}.

\subsection{Type Model}

To ensure the accuracy and completeness of the object's structural information at different locations in the program, 
we introduced structure-flow-sensitive analysis for objects, drawing from flow-sensitive pointer analysis. 
Our structure-flow-sensitive model only allows weak updates to the object's type and does not permit strong updates. 
This means that once an object has a certain type, this type remains intact throughout the object's lifecycle. 
As a result, the type set of an object accumulates monotonically. 
We adopted this approach because existing pointers of various types pointing to an object will not disappear due to changes in the object's type. 
These pointers can still access fields of the object's previous type. 
Therefore, when the type of an object changes, we cannot discard the existing type of the object. 
Moreover, since our structure-flow-sensitivity is based on flow-sensitive pointer analysis, the object possesses different structural information at different program points. 
When the type of an object changes, we assume an undeclared union relationship exists between the new type and the object's existing types. 
For instance, in Figure \ref{fig:motivationA}, we interpret the $T1$ and $T2$ types of object $o$ as an undeclared union type. 
We use $point$-$to$-$types$ ($pt.t$) to record the various types an object possesses at specific locations in the program. 
For example, $pt.t(\ell,o)$ represents the set of types that object $o$ possesses at location $\ell$.

Due to our introduction of structure-flow-sensitivity for objects, our type model permits the conversion of object types. 
What we need to emphasize here is that we only focus on the conversion of the object's structural types (types that encompass more than one element, such as arrays, structs, etc.), 
as MTO-SS enhances the accuracy of the points-to analysis results through the structural information of the object. 
The conversion of an object's structure can be categorized into two scenarios. 
One is to directly interpret the object as another type. 
This situation is widespread in C/C++, and some C++ functions also have the same effect, such as static\_cast. 
Another case is the $dynamic\_cast$ type conversion function unique to C++, which checks the legality of the conversion by inspecting the relationship between the target type and the initial type of the object on the inheritance tree. 
This dynamic type conversion method is very intelligent. 
The pointer after this type conversion always points to the memory location where the target class of the object resides. 
For example, in Figure \ref{fig:motivationB}, in the type conversion statement $R* ri=dynamic\_cast<R*>(baseL)$, the pointer $ri$ will point to the location of the $struct$ $R$ within the object $o1$.

We not only consider type conversion from the base address of an object, but also type conversion of sub-objects, that is, interpreting a part of the object as another type. 
When converting the type of a sub-object, we also need to consider its offset relative to the object's base address, 
which allows for a more accurate representation of the type information for each part of the object. 
This is always feasible, as the offset of the sub-object relative to the object's base address is always easily obtainable. 


As in previous work, we consider that a pointer pointing to the first field of a structure also points to the entire object, 
and vice versa, a pointer pointing to an object is also considered to point to the first field of the structure, this is equivalent relationships. 
For example, in Figure \ref{fig:nestStruct}, the pointer points to the first field $long$ $y$ of object $o$, which is equivalent to the pointer pointing to the object $o$ itself. 
For cases where the structure contains nested structures, we unfold the nested structure until it is expanded to basic elements. 
As shown in Figure \ref{fig:nestStruct}, we unfold the struct $A$ inside struct $B$. 
After unfolding, struct $B$ contains three basic elements: $y$, $x$, and $v$.


\begin{figure}[htbp]
\raggedright
\begin{lstlisting}[numbers=left, numberstyle=\normalfont, numbersep=-8pt, commentstyle=\color{green!60!black}, language=C++]
    typedef struct {
        int x; 
        int v;
    }A;
    typedef struct { 
        long y; 
        A a;
    }B;
    int main(){
        B* p = (B*)malloc(sizeof(B)); //o
        void *q = &(p->a); //o1
    }
\end{lstlisting}
    
\caption{A nested structure example.}
\label{fig:nestStruct}
\end{figure}

\begin{figure*}
  \centering

  \begin{subfigure}{1.0\textwidth}
    \begin{align*}
      \begin{array}{l}
        \textbf{{\scriptsize [STACK/GLOBAL]}} \\
        $$ \ \ \ \ell  :   p = alloca \ t,nbytes$$   \\
        \hline
        $$o    \in    pt(\ell,p)   \ \  t   \in    pt.t(\ell,o)$$
      \end{array} \ & \quad \quad \quad
      \begin{array}{l}
        \textbf{{\scriptsize [HEAP]}} \\
        $$\ell \ : \  p = malloc \ nbytes$$   \\
        \hline
        \quad \quad  $$o    \in    pt(\ell,p) $$ 
      \end{array} \ & \quad
      \begin{array}{l}
        \textbf{{\scriptsize [CAST]}} \\
        $$\ell :  p = (t) q \ \ \ \ \ell'  \xrightarrow{ q }  \ell  \ \ \ \ \ell''  \xrightarrow{ o }  \ell  \ \ \ \  o    \in    pt(\ell',q) $$   \\
        \hline
         $$pt(\ell',q) \subseteq pt(\ell,p)    \ \   \{t\} \cup pt.t(\ell',o )   \subseteq    pt.t(\ell,o)$$
      \end{array}
      \end{align*}
  \end{subfigure}  


  \begin{subfigure}{1.0\textwidth}
    \begin{align*}
      \begin{array}{l}
        \textbf{{\scriptsize [COPY]}} \\
        $$\ell :   p = q \ \ \  \ell'  \xrightarrow{ q }  \ell \ \ \  \ell''  \xrightarrow{ o }  \ell \ \ \   o  \in  pt(\ell',q) $$    \\
        \hline
        $$pt(\ell',q) \subseteq pt(\ell,p) \ \ pt.t(\ell',o )   \subseteq    pt.t(\ell,o) $$ 
      \end{array} \ & \ \ \ \ \ \ \
      \begin{array}{l}
        \textbf{{\scriptsize [CONSTRUCTOR-C++]}} \\
        $$ \ell \xrightarrow{calls}  constructor(t \ \ o,...)     $$  \\
        \hline
        $$ \ \ \ \ \ \ \ \ t \in pt.initT(o)   $$
      \end{array} \ & \ \
      \begin{array}{l}
        \textbf{{\scriptsize [PHI]}} \\
        $$\ell  :   p = \phi(q,x) \ \  \ell'  \xrightarrow{ q }  \ell  \ \  \ell''  \xrightarrow{ o }  \ell   \\
        o \in pt(\ell',q) \ \  \ell'''  \xrightarrow{ x }  \ell \ \  \ell''''  \xrightarrow{ o' }  \ell \ \ o' \in pt(\ell''',x) $$     \\
        \hline
        $$  pt.t(\ell''',o' )   \subseteq    pt.t(\ell,o')  \ \ pt.t(\ell',o )   \subseteq    pt.t(\ell,o) \\ 
        pt(\ell',q) \cup  pt(\ell''',x)  \subseteq    pt(\ell,p) $$ 
      \end{array}
      \end{align*}
    
  \end{subfigure}

  \begin{subfigure}{1.0\textwidth}
    \begin{align*}
      \begin{array}{l}
        \textbf{{\scriptsize [LOAD]}} \\
        $$\ell  :   p = *q \ \ \  \ell''  \xrightarrow{ q }  \ell  \ \ \  \ell'  \xrightarrow{ o }  \ell   \ \ \  o    \in    pt(\ell'',q)$$   \\
        \hline
         $$pt(\ell',o)    \subseteq    pt(\ell,p)  \ \ pt.t(\ell'',o )   \subseteq    pt.t(\ell,o) $$ 
      \end{array} \ & \ \ \ \ \ \ \ \ \ \
      \begin{array}{l}
        \textbf{{\scriptsize [STORE]}} \\
        $$\ell  :  *p = q \ \ \ \  \ell'  \xrightarrow{ q } \ell  \ \ \ \ \ell''  \xrightarrow{ p }  \ell \ \ \ \  \ell'''  \xrightarrow{ o }  \ell  \\
        o    \in    pt(\ell'',p) \ \ \ \ \  \ell''''  \xrightarrow{ o' }  \ell  \ \ \ \ \ o'   \in    pt(\ell',q) $$   \\
        \hline
        $$pt(\ell',q)    \subseteq    pt(\ell,o) \ \ pt.t(\ell',o' )   \subseteq    pt.t(\ell,o') $$ 
      \end{array} \ & \ \
      \begin{array}{l}
        \textbf{{\scriptsize [STRONG-UPDATE]}} \\
        \ell  :   *p = q \ \  \ell'  \xrightarrow{ p }  \ell  \ \  \ell''  \xrightarrow{ q }  \ell \\
        \ \ \ell'''  \xrightarrow{ o }  \ell \ \ \ \quad o \in pt(\ell',p)   \\
        \hline
        pt(\ell,o)  = Update(\ell,p,q,o)   
      \end{array}
      \end{align*}
    
  \end{subfigure}

  \begin{subfigure}{1.0\textwidth}
    \centering
    \begin{align*}
      \begin{array}{cc}
        Update(\ell,p,q,o) = \left\{
        \begin{array}{ll}
            pt(\ell,q) & \text{if } pt(\ell,p) \equiv \{o\} \land o \in \text{singletons} \\
            pt(\ell,o) & \text{else if } pt(\ell,p) \equiv \varnothing \\
            pt(\ell,o) \cup pt(\ell,q) & \text{otherwise}
        \end{array}
        \right.
        \ & \ \ \ \ 
        \begin{array}{l}
          \textbf{{\scriptsize [DYNAMIC\_CAST-C++]}} \\
          $$\ell  :   p = dynamic\_cast\langle t*\rangle (q) \quad \ \ \ \ \ell'  \xrightarrow{ q }  \ell    $$   \\
          \hline
          $$\forall o_i \in pt(\ell',q) \land t' \in pt.initT(o_{i\_base})  \land t \in t' ,  \\
          off = offset(t' \rightarrow t), (o_{i\_base} \rightarrow off) \in     pt(\ell,p)$$ 
        \end{array}
      \end{array}
    \end{align*}
  \end{subfigure}

  \begin{subfigure}{1.0\textwidth}
    \begin{align*}
      \begin{array}{l}
        \textbf{{\scriptsize [ARRAY-CONST]}} \\
        $$\ell  :   p = \&q[c] \ \ \ \ \ \ell'  \xrightarrow{ q }  \ell \ \ \ \ \ off = offset(q[c])  $$ \\
        \hline
        $$ \forall  o_i \in pt(\ell',q) \land t \in pt.t(\ell',o_i) \land off \in pt.o(t) ,   \\
        (o_i \rightarrow off)  \in pt(\ell,p) $$
      \end{array} \ & \ \ \qquad  \qquad \qquad \   
      \begin{array}{l}
        \textbf{{\scriptsize [ARRAY-VAR]}} \\
        $$ \ \ \ \ \ \ \ell  :   p = \&q[j]  \ \ \  \ \ \ \ \ \ \ \  \ell'  \xrightarrow{ q }  \ell    $$ \\
        \hline
         $$ \forall  o_i \in pt(\ell',q) \land t \in pt.t(\ell',o_i) \land \forall off \in pt.o(t) ,  \\
          (o \rightarrow off)  \in pt(\ell,p) $$   
      \end{array}
      \end{align*}
  \end{subfigure}

  \begin{subfigure}{1.0\textwidth}
    \begin{align*}
      \begin{array}{l}
        \textbf{{\scriptsize [FIELD]}} \\
        $$\ell  :   p = \&q \rightarrow fld   \ \   \ell'  \xrightarrow{ q }  \ell \ \ off = offset(q \rightarrow fld ) $$ \\
        \hline
         $$\forall  o_i \in pt(\ell',q) \land t \in pt.t(\ell',o_i) \land off \in pt.o(t)  , \\
         (o_i \rightarrow off) \in pt(\ell,p)  $$  
      \end{array} \ & \ \ \qquad   \qquad \ \ \ 
      \begin{array}{l}
        \textbf{{\scriptsize [FIELD-VAR]}} \\
        $$\ell  :   p = \&q \rightarrow fld    \ \ \  \ \ \ \ \ \ \ \   \ell'  \xrightarrow{ q }  \ell  $$ \\
        \hline
         $$\forall  o_i \in pt(\ell',q) \land t \in pt.t(\ell',o_i) \land \forall off \in pt.o(t)  , \\
         (o_i \rightarrow off) \in pt(\ell,p)  $$  
      \end{array}
      \end{align*}
  \end{subfigure}

  \begin{subfigure}{1.0\textwidth}
    \begin{align*}
      \begin{array}{l}
        \textbf{{\scriptsize [CALL]}} \\
        $$\ell  :   p = q(...,a,...)  \ \  \ \mu(o) \ \  \ \ell' : m'(...,a'...)  \ \ \ o = \chi(o)      $$   \\
        $$ \ell''  \xrightarrow{ q }  \ell  \ \ \ \ \ \ \ o_{m'} \in pt(\ell'',q) \ \ \ \ \ \ \ \ell^* \xrightarrow{ a } \ell  \  \ \ \ \ \ \ \ell^* \xrightarrow{ o } \ell $$ \\
        \hline
        $$     pt(\ell^*,a) \subseteq pt(\ell',a') \ \ pt(\ell^*,o) \subseteq pt(\ell',o) \\
        pt.t(\ell^*,o) \subseteq pt.t(\ell',o) $$ \\
      \end{array}  \ & \ \qquad  \qquad  \qquad    \ \ \ \ \ \
      \begin{array}{l}
        \textbf{{\scriptsize [RET]}} \\
        $$\ell  :   p = q(...) \ \ \ o = \chi(o)  \ \ \     \ell'' : ret \ p' \ \ \ \mu(o)   \\
        \ell'  \xrightarrow{ q }  \ell \ \ \ o_{m} \in pt(\ell',q) \ \ \ \ell^*  \xrightarrow{ o }  \ell \ \ \  \ell^*  \xrightarrow{ p' }  \ell $$  \\
        \hline
        $$ pt(\ell^*,p') \subseteq pt(\ell,p) \ \ pt.t(\ell^*,o )   \subseteq    pt.t(\ell,o) $$
      \end{array}
      \end{align*}
  \end{subfigure}

  \vspace{0.3cm}
 
  \begin{subfigure}{0.45\textwidth}
    $pt.t(\ell,o) $ : $\mathcal{L \times A} \mapsto 2^{\mathcal{T}}$  \ \ \  $o$'s type set after $\ell$
  \end{subfigure}
  \begin{subfigure}{0.45\textwidth}
    $o \rightarrow off \ \  \overset{\cdot}{\sim} \ \ \widehat{o_{off}}  $  \ \ \ \ \ \ $\widehat{o_{off}}$ represent sub-object.
  \end{subfigure}

  \vspace{0.3cm}

  \begin{subfigure}{0.45\textwidth}
    $\ell  \xrightarrow{ v }  \ell' $ :  $\mathcal{L \times V \times L} $  \ \ \   $v$'s value flow
  \end{subfigure}
  \begin{subfigure}{0.45\textwidth}
    $pt(\ell,v)  $ : $\mathcal{L \times V} \mapsto 2^{\mathcal{A}}$  \ \ \  $v$'s points-to set after $\ell$ 
  \end{subfigure}

  \vspace{0.3cm}

  \begin{subfigure}{1.0\textwidth}
   \ \ \ \ \ \ \ \ \ \ $off = offset(o \rightarrow fld) $ : $offset(\mathcal{O \times F}) \mapsto off$ \ \ \   get the $fld$ offset relative to object base address
  \end{subfigure}

  \caption{MTO-SS's inference rules}
  \label{fig:inferenceFig}
  
\end{figure*}

\subsection{Structure-flow-sensitive Analysis}

\subsubsection{sub-object representation}

Before introducing our Structure-flow-sensitive pointer analysis, we first explain how to represent the different fields of an object, or in other words, how to represent a part of an object. 
We accurately represent the offset position of the object that the pointer points to. 
We cannot represent the pointer as pointing to a field of the object because an object may have multiple types. 
To accelerate points-to analysis and provide a more straightforward representation, we create sub-objects for the offset of each field of an object, recording the offset of the sub-object relative to the object's base address in the sub-object. 
These sub-objects are constructed on demand based on the memory offsets of the objects accessed by the pointers. 
This means the memory offset of each field relative to the starting address of the object signifies a distinct sub-object. 
For example, in the [FIELD] rule, $o \rightarrow off$ is represented as $\widehat{o_{off}}$, and these two symbols are equivalent. 
Put simply, if two distinct fields of the same object have the same memory offset relative to the object, then these two fields will be represented by the same sub-object. 
This approach indirectly addresses the issue of field equivalency among multiple structures within the same object. 
For example, in Figure \ref{fig:motivationA}, the object $o$ has both type $T1$ and type $T2$, so the fields $s1$ and $st$ are equivalent fields. 
In our method, these two fields will be represented by the same sub-object. 
Our approach effectively addresses the issue of field equivalence. 

Another advantage of using sub-objects to represent different offsets of an object is that sub-objects can be handled in the same manner as basic objects. 
The analysis of sub-objects can employ the same inference rules as the analysis of basic objects. 
Similarly, the analysis of sub-object type sets also follows the same inference rules as the analysis of basic object types. 
Thus, the objects in our inference rules include both basic objects and sub-objects. 
However, there is a particular point that needs special mention. 
In the [DYNAMIC\_CAST-C++], [ARRAY-CONST], [ARRAY-VAR], [FIELD-VAR], and [FIELD] rules, sub-objects are created and inherit the type information from the corresponding offset of their base object. 
Although this feature is not depicted in the inference rules for the sake of simplicity, it is considered during our pointer analysis process. 
For example, in Figure \ref{fig:nestStruct}, the pointer $q$ points to the location of the substructure $A$ $a$ within object $o$. 
Here, we create sub-object $o1$ of object $o$. 
Sub-object $o1$ inherits the substructure $A$ from object $o$, which means that sub-object $o1$ has the type $struct$ $A$ at this position. 

\subsubsection{Initial Types of Objects}

[STK/GLOBAL] and [HEAP] are two inference rules used to analyze the initial types of stack, global, and heap objects. 
How these two inference rules handle objects is the same as the basic flow-sensitive analysis method. 
The difference lies in the fact that [STK/GLOBAL] can retrieve the type of the object from the allocation statement. 
Therefore, we add type $t$ to the set $pt.t(\ell,o)$ to indicate that object $o$ has acquired a new type $t$ at program location $\ell$. 
On the other hand, [HEAP] cannot obtain the object's type from the allocation statement, so its $pt.t(\ell,o)$ target set is left empty and is not recorded. 

\subsubsection{Propagation of Object Types}

The six rules [COPY], [PHI], [LOAD], [STORE], [CALL], and [RET] are only involved in the propagation of objects and object structure sets. 
These rules do not involve adding elements to the object structure set, 
nor do they improve the accuracy of pointer analysis through the structural information of the object. 
The way these rules propagate objects is the same as in flow-sensitive analysis. 
For example, both the [COPY] and [PHI] rules are simple assignment instructions, propagating the points-to information of the right-hand pointer to the left side. 
Specifically, the [PHI] rule merges the points-to sets of the two pointers on the right and propagates them to the left. 
However, the distinction between our method and basic flow-sensitive analysis is that our inference rules also ensure the propagation of the object type set. 
When an object is propagated, its type set is also propagated to the current location in a flow-sensitive manner. 
This means the set of object types is copied to the current position without altering the object type set itself. 
However, the location of the source object's type set is different from the location where the object is defined. 
The source location of the object type set is the same as the location of the source pointer. 
For example, in the [COPY] rule, both the source location of the type set $pt.t(\ell',o)$ for object $o$ and the location of the source pointer $q$ originate from $\ell'$. 
The reason for this is that the conversion of the object's type always acts directly on the pointer and indirectly on the object. 
The specific method of propagating object structures is shown in Figure \ref{fig:inferenceFig} in the [COPY] and [PHI] inference rules.

The treatment of object propagation in [LOAD] and [STORE] is no different from the standard flow-sensitive method. 
However, it's worth noting that the object type set is also propagated alongside the object to the current location. 
Similar to the [COPY] and [PHI] rules, the type set of the object comes from the same location as the pointer, rather than from the location where the object is defined. 
The [STRONG-UPDATE] rule implements the standard object update method in flow-sensitivity \cite{hardekopf2009semi, lhotak2011points}. 
For the store instruction $*p=q$, when the pointer $p$ points to a single object in any circumstance, a strong update is performed, that is, 
it eliminates the original content pointed to by $*p$, and makes $*p$ point to the content pointed to by $q$. 
This is ensured by checking all def-use chains reaching the store instruction to ensure that $p$ only points to one object. 
When the pointer $p$ points to multiple objects, a weak update is performed, merging the points-to set of pointer $q$ with the object set pointed to by $*p$. 
When the pointing set of the pointer $p$ is empty, nothing is done. 
The inference rules are displayed in Figure \ref{fig:inferenceFig}.

The inter-procedural points-to propagation is accomplished through the [CALL] and [RET] rules. 
MTO-SS employs an on-the-fly approach for resolving indirect function calls, allowing for a more accurate identification of the called functions during the points-to analysis. 
The points-to variables propagate from the actual parameters at the call site to the corresponding formal parameters of the callee. 
The points-to information for address-taken variables is propagated through indirect value-flows as in section \ref{sec:valueflow}. 
The structure information of the object is also propagated along with the object, as illustrated by the rule in Figure \ref{fig:inferenceFig}. 

\subsubsection{Type Conversion for Objects}

The two inference rules [CAST] and [DYNAMIC\_CAST-C++] are used to handle the type conversion of objects. 
We first explain the static type conversion of an object, which means interpreting the object according to another type without changing the location the pointer points to. 
This situation is handled by the [CAST] rule. 
The [CAST] rule handles objects in the same way as the [COPY] rule. 
The difference with the [COPY] rule lies in the [CAST] rule's handling of object types. 
The [CAST] rule adds the converted type to the type set of the object at the current program location. 
In the [CAST] rule, we merge the converted type $t$ with the original type set $pt.t(\ell',o)$ of the object and add it to the object's current location type set $pt.t(\ell,o)$. 
It should be specially noted here that, since our method is structure-sensitivity, we only focus on the conversion of structure types. 
Therefore, in the MTO-SS inference rules, $t$ represents a structure. 
We need to clarify the cases of type conversion for a part of an object. 
In our inference rules, there is no mention of type conversion for a part of an object. 
However, our method ingeniously addresses this issue. 
Since we represent different fields of an object as distinct sub-objects based on their offset in memory, we can also handle type conversions of sub-objects in the same way as we handle basic objects. 
The [CAST] rule can analyze both base object type conversions and sub-object type conversions. 


\begin{figure}[htbp]
\raggedright
\begin{lstlisting}[numbers=left, numberstyle=\normalfont, numbersep=-8pt, commentstyle=\color{green!60!black}, language=C++]
    class B {virtual void f() { } };
    class D:B {virtual void f() { } };
    class UB {virtual void f() { } };
    class UD:UB {virtual void f() { } };
    int main(){
      D* derived = new D(); //o
      B* base = derived;
      UB* uB = (UB*)base;
      UD* uD=dynamic_cast<UD*>(uB);//fail   
    }
\end{lstlisting}
      
  \caption{A $dynamic\_cast$ conversion fails example }
  \label{fig:dynamicUnRelate}
  \end{figure}



The [DYNAMIC\_CAST-C++] and [CONSTRUCTOR-C++] inference rules are used for dynamic type conversions caused by C++'s $dynamic\_cast$. 
The $dynamic\_cast$ type cast function differs from static type casting. 
The $dynamic\_$$cast$ function alters the offset of the object the pointer points to, based on the relationship between the conversion type and the object's initial type. 
If there is no inheritance relationship between the conversion type and the object's initial type, the type conversion will fail and return a null pointer. 
As shown in Figure \ref{fig:dynamicUnRelate}, using the $dynamic\_cast$ function to convert object $o$ to class $UD$ results in a failed conversion, returning a null pointer. 
The $dynamic\_cast$ function determines the legitimacy of type conversions based solely on the type initialized in the constructor. 
To obtain the initial type of a C++ object, we have established the [CONSTRUCTOR-C++] rule. 
The first parameter of the C++ constructor is the 'this' pointer, which always points to the object being initialized and contains the object's class information. 
The utility of the initial type for C++ objects is limited only to objects containing virtual functions. 
For objects that do not contain virtual functions, the initial type is of no use since these objects will not execute type conversion through $dynamic\_cast$. 
Hence, we only record the initial type information for objects containing virtual functions. 
We use $point$-$to$-$initType$ ($pt.initT$) to represent the initial type of the C++ object. 
The recorded initial type is global for the object and stays unchanged until the constructor is invoked for that object again.

To handle dynamic type conversion with the $dynamic\_cast$ function, we set up the [DYNAMIC\_CAST-C++] rule. 
Unlike type conversion for the entire object or sub-object conversion, $dynamic\_cast$ does not add new types to the object's type set. 
This is because $dynamic\_cast$ always performs type conversions between the initial type of the object and its base class, which means the target class is always a subtype of the initial class. 
For example, in Figure \ref{fig:motivationB}, the object $o1$ can only be converted to $B$, $L$, $R$, or $D$ through $dynamic\_cast$. 
However, for objects converted through $dynamic\_cast$, the location of the object that the pointer points to changes, which means the pointer may point to a different sub-object. 
For instance, in the statement $R* ri=dynamic\_cast<R*>(baseL)$ in Figure \ref{fig:motivationB}, the pointer $ri$ points to the location of the subtype $struct \ \ R$ of object $o1$. 
In the [DYNAMIC\_CAST-C++] rule, we examine the relationship between the target type and the initial type of the object which is recorded in $pt.initT$. 
This examination determines the position of the target type within the initial type, as well as the position that the resulting pointer points to. 
What is being checked here is the basic object's $pt.initT$ set, which is the $o_{base}$ in the inference formula. 
The reason for this is that the target of a $dynamic\_cast$ conversion statement may be a sub-object. 
It does not make sense to perform $dynamic\_cast$ conversions based on the sub-object's $pt.initT$ set.

\subsubsection{Structure-sensitive Pointer Analysis for Objects}

The [ARRAY-CONST], [ARRAY-VAR], [FIELD-VAR] and [FIELD] rules are used to infer object field access. 
In our method, both the [ARRAY-CONST] and [FIELD] rules utilize the structural information of the object to improve pointer analysis accuracy. 
MTO-SS conducts a pre-analysis of the program's structure and determines the offsets of all fields within it by expanding all internal substructures to basic elements. 
This is done to check the legitimacy of access during the pointer analysis process in combination with the object's structural information. 
MTO-SS employs a set, denoted as $point$-$to$-$offsets$ ($pt.o$), to represent the offsets of all basic elements inside a structure. 
For instance, by unfolding all complex types within the structure $t$ to basic elements, the offsets of these basic elements in memory are denoted by $pt.o(t)$. 
As an example, in Figure \ref{fig:nestStruct}, after unfolding the struct $A$ inside struct $B$, $pt.o(B)$ will contain offsets \{0, 8, 12\}. 
It's worth noting that field offsets take into account the memory size and alignment of the type, adhering to the alignment method of the C/C++ standard. 
For example, in a 64-bit operating system, the 'int' type occupies 4 bytes and aligns with 4 bytes, while the 'long' type occupies 8 bytes and aligns with 8 bytes.

The [ARRAY-CONST] and [FIELD] rules both use a similar method to filter out spurious points-to relations. 
Both rules check whether each type within the object's type set contains the field offset accessed by the current statement. 
If the field offset isn't found, that points-to relation is filtered out. 
If the field offset is present, a sub-object for that offset will be created, and the pointer will point to the created sub-object. 
[ARRAY-VAR] and [FIELD-VAR] describe the scenarios of accessing array elements and structure elements through a variable. 
In such cases, we conservatively assume that the pointer might point to any sub-object of the object. 
The inference rules are already shown in Figure \ref{fig:inferenceFig}. 


\subsection{Soundness}

Our method is based on flow-sensitive analysis and further filters spurious points-to relationships by incorporating the object's structural information, 
making the results of pointer analysis more accurate. 
For cases that do not contain structural information and access allocated memory blocks directly as basic types, 
such as i8*, we will not use the structure to filter the pointer relationships. 
Therefore, as long as the fully sparse flow-sensitive pointer analysis is sound, our method is also sound.

\section{Evaluation}

\begin{table*}[htbp]
  \caption{The first column indicates the size of the compiled bitcode, 
  the second column presents the number of lines of code, 
  columns 3 through 5 display the percentage of no-alias query results relative to the total query results for SPARSE, TYPECLONE, and MTO-SS, 
  columns 6 and 7 highlight the accuracy improvement of MTO-SS compared to SPARSE and TYPECLONE, 
  column 8 shows the additional time overhead introduced by MTO-SS.}
  \label{tab:aliasRes}
  \begin{tabular}{lc@{\hspace{1.5em}}c@{\hspace{1.5em}}cccccc}
  \hline
  \multirow{2}{*}{Benchmark} & \multirow{2}{*}{Size} & \multirow{2}{*}{LOC} & SPARSE          & TYPECLONE       & \multicolumn{4}{c}{MTO-SS}   
                                                                                                                                                                                \\ \cmidrule(r){4-4} \cmidrule(lr){5-5} \cmidrule(l){6-9}
                             &                       &                      & No-alias result & No-alias result & No-alias result & \begin{tabular}[c]{@{}c@{}}Improve-\\ SPARSE\end{tabular} & \begin{tabular}[c]{@{}c@{}}Improve-\\ TYPECLONE\end{tabular} & \begin{tabular}[c]{@{}c@{}}Extra \\ running time\end{tabular} \\ \hline
  du                         & 455 KB                & 21811                & 72.82\%         & 97.50\%         & 96.61\%         & 32.67\%                                                   & -0.91\%                                                      & 0.935s                                                        \\
  date                       & 489 KB                & 26151                & 73.57\%         & 93.38\%         & 98.78\%         & 34.27\%                                                   & 5.78\%                                                       & 0.2s                                                          \\
  touch                      & 423 KB                & 22717                & 73.19\%         & 93.29\%         & 98.61\%         & 34.73\%                                                   & 5.70\%                                                       & 0.183s                                                        \\
  ptx                        & 187 KB                & 8547                 & 82.83\%         & 96.26\%         & 99.17\%         & 19.73\%                                                   & 3.02\%                                                       & 0.042s                                                        \\
  csplit                     & 179 KB                & 7375                 & 78.65\%         & 97.52\%         & 98.83\%         & 25.66\%                                                   & 1.34\%                                                       & 0.02s                                                         \\
  expr                       & 259 KB                & 12676                & 84.48\%         & 95.97\%         & 98.43\%         & 16.51\%                                                   & 2.56\%                                                       & 0.066s                                                        \\
  tac                        & 127 KB                & 4880                 & 80.59\%         & 95.22\%         & 98.46\%         & 22.17\%                                                   & 3.40\%                                                       & 0.016s                                                        \\
  nl                         & 131 KB                & 4892                 & 80.02\%         & 95.24\%         & 97.99\%         & 22.46\%                                                   & 2.89\%                                                       & 0.009s                                                        \\
  mv                         & 506 KB                & 23505                & 79.40\%         & 93.97\%         & 96.82\%         & 21.94\%                                                   & 3.03\%                                                       & 0.778s                                                        \\
  ls                         & 570 KB                & 25851                & 85.00\%         & 95.17\%         & 99.38\%         & 16.92\%                                                   & 4.42\%                                                       & 0.159s                                                        \\
  ginstall                   & 559 KB                & 27742                & 94.65\%         & 97.14\%         & 99.16\%         & 4.76\%                                                    & 2.08\%                                                       & 0.226s                                                        \\
  sort                       & 499 KB                & 24848                & 80.30\%         & 93.97\%         & 99.14\%         & 23.46\%                                                   & 5.50\%                                                       & 0.168s                                                        \\ \hline
  \multicolumn{6}{l}{Average}                                                                                                     & 22.94\%                                                   & 3.23\%                                                       & 0.2235s                                                       \\ \hline
  \end{tabular}
  \end{table*}

The goal of our experiments is to compare how much the accuracy of MTO-SS improves compared to SPARSE \cite{hardekopf2011flow} and TYPECLONE \cite{barbar2020flow}, and to evaluate the additional performance overhead introduced by our method. 
Moreover, through the experiment, we once again demonstrate that it is common for objects to have multiple types throughout their lifecycle. 
We implemented MTO-SS based on LLVM-13 \cite{llvm@website} and SVF-2.5 \cite{svf@website}, and the source code is publicly available at \url{https://github.com/Frankenstein-bit/Structure-Flow-Sensitive}. 
The structural information we used comes directly from the intermediate representation (LLVM IR) generated after the LLVM compilation of the source code. 
All experiments were conducted on a machine running 64-bit Ubuntu 20.04 LTS, equipped with a 13th Gen Intel(R) Core(TM) i7-13700H processor at 2.640GHz and 32GB of RAM.

\begin{table*}[htbp]
  \caption{Static Structural Data on Benchmarks. 
  The first column indicates the number of type cast instructions in each program, 
  the second column shows the number of structures (including array types) in each program, 
  the third column presents the maximum number of elements a structure can contain, 
  the fourth column represents the number of basic objects, 
  the fifth column indicates the maximum number of structures a single object can possess, 
  the sixth column shows the number of basic objects that have multiple structures.}
  \label{tab:staticmto}
  \begin{tabular}{lcccccc}
  \hline
  Benchmark & Cast Instructions & Structure Count & \begin{tabular}[c]{@{}c@{}}Max Elements \\ in a Structure\end{tabular} & \begin{tabular}[c]{@{}c@{}}Basic Object \\ Count\end{tabular} & \begin{tabular}[c]{@{}c@{}}Basic Object \\ with Most Structures\end{tabular} & \begin{tabular}[c]{@{}c@{}}Multi-structure Basic \\ Object Count\end{tabular} \\ \hline
  du        & 942               & 132          & 1024                                                                & 1971                                                          & 6                                                                         & 31                                                                         \\
  date      & 1595              & 152          & 2000                                                                & 1689                                                          & 6                                                                         & 24                                                                         \\
  touch     & 1477              & 129          & 2000                                                                & 1443                                                          & 6                                                                         & 31                                                                         \\
  ptx       & 485               & 77           & 418                                                                 & 723                                                           & 4                                                                         & 7                                                                          \\
  csplit    & 299               & 84           & 438                                                                 & 789                                                           & 2                                                                         & 8                                                                          \\
  expr      & 741               & 92           & 310                                                                 & 1187                                                          & 4                                                                         & 8                                                                          \\
  tac       & 184               & 63           & 8192                                                                & 552                                                           & 2                                                                         & 5                                                                          \\
  nl        & 185               & 68           & 443                                                                 & 532                                                           & 2                                                                         & 4                                                                          \\
  mv        & 1043              & 139          & 1024                                                                & 2400                                                          & 6                                                                         & 38                                                                         \\
  ls        & 1427              & 180          & 8192                                                                & 2555                                                          & 9                                                                         & 42                                                                         \\
  ginstall  & 1352              & 137          & 4096                                                                & 2588                                                          & 4                                                                         & 30                                                                         \\
  sort      & 1213              & 149          & 4096                                                                & 2081                                                          & 5                                                                         & 13                                                                         \\ \hline
  \end{tabular}
  \end{table*}

The reason we compare our method with SPARSE \cite{hardekopf2011flow} and TYPECLONE \cite{barbar2020flow} in our experiments is: 
SPARSE and TYPECLONE are the only LLVM-based, whole-program, flow-sensitive pointer analysis methods available for C/C++. 
It's worth noting that TYPECLONE also considers object structural information. 
However, it only allows an object to have a single type throughout its lifecycle. 
We employ the 12 largest programs from GNU Coreutils 8.31 \cite{gnu@website} to evaluate the accuracy and analysis time of our method. 
To test accuracy, we perform alias queries on pointers. 
If the points-to sets of two pointers intersect, the two pointers are considered to be aliases. 
If there's no intersection in the points-to sets of two pointers, this indicates a no-alias relationship between them. 
The no-alias results in alias queries indicate the accuracy of pointer analysis. 
Therefore, the higher the proportion of no-alias results in the total number of alias queries, the more accurate the pointer analysis results are. 
Thus, we use the percentage of no-alias results in the total number of alias queries to represent the accuracy of pointer analysis. 
The percentage of MTO-SS no-alias results in the total number of alias queries is shown in Table \ref{tab:aliasRes}, 
and the percentage of SPARSE no-alias results in the total number of alias queries is also displayed in Table \ref{tab:aliasRes}. 
We found that for all benchmark projects, MTO-SS significantly improves accuracy over SPARSE. 
MTO-SS only has a 4\% improvement over SPARSE in the "ginstall" project, due to SPARSE's already high analysis accuracy for this project. 
Overall, MTO-SS improves the accuracy by an average of 22.94\% compared to SPARSE. 
Furthermore, we also compared the analysis results of MTO-SS with TYPECLONE. 
TYPECLONE is also flow-sensitivity and takes into account the type information of objects. 
However, TYPECLONE assumes that an object has only one type throughout its entire lifecycle. 
In theory, TYPECLONE should produce a higher percentage of no-alias results, with some of these results likely resulting from pointer analysis errors. 
However, our approach still makes a higher percentage of no-alias query results than TYPECLONE while ensuring soundness. 
Only in the "du" project, the accuracy of MTO-SS is 0.91\% lower than TYPECLONE. 
Across all projects, the accuracy of MTO-SS has increased by an average of 3.23\% compared to TYPECLONE. 

We also analyzed the additional time introduced by MTO-SS. 
The analysis time here only includes the time required for our multi-structure object pointer analysis methods and does not include the pre-analysis time before flow-sensitive analysis. 
The analysis time is displayed in Table \ref{tab:aliasRes}. 
It can be seen that the additional time overhead introduced by MTO-SS is very low, with an average extra time overhead of only 0.2235 seconds. 
Our method does not produce significant additional time overhead, and introducing our method into flow-sensitive pointer analysis will not have a major impact on performance.

We also analyzed static data related to structures and type conversions in the program, as shown in Table \ref{tab:staticmto}. 
This data includes the number of type conversion instructions, 
the number of structures (including arrays, structs, etc.), 
and the number of elements in the largest structure, where the element count reflects the number after unfolding internal substructures. 
Additionally, we counted the number of basic objects. 
Here, only the count of basic objects is taken into consideration, as they more accurately reflect the number of structures a basic object contains. 
We also documented the maximum number of structures a single object possesses within each program. 
At the same time, we noted another significant data point: how many basic objects in each program have more than one structure. 
From the last row of the table, it can be seen that it's quite common for objects to have multiple structures. 
Therefore, we can conclude that in every program there exists a certain number of basic objects with multiple structures.

Overall, through experiments, we demonstrated that incorporating structural information of objects can greatly enhance the accuracy of pointer analysis. 
Moreover, we have demonstrated through experiments that introducing MTO-SS into flow-sensitive analysis does not have a significant impact on performance. 
Furthermore, our experiments also showed that it is common for objects to possess multiple structures.

\section{Related work}

Researchers have extensively studied whole-program flow-sensitive pointer analysis. 
The foundation of traditional flow-sensitive analysis is built upon the iterative data-flow framework \cite{choi1993efficient, emami1994context, kam1977monotone, wilson1995efficient}. 
This framework models the program as a control flow graph(CFG), where nodes represent statements and edges signify control-flow. 
By repeatedly propagating the data-flow on the CFG, it eventually reaches a fixed point. 
While this method yields fairly precise points-to results, it also propagates a lot of useless points-to information in each iteration. 
This results in consuming too much memory and runtime to maintain unnecessary points-to information, which makes it impractical for large-scale programs. 
In order to eliminate the propagation of unnecessary points-to information during flow-sensitive analysis, 
researchers proposed sparse flow-sensitive pointer analysis method \cite{hardekopf2009semi, hardekopf2011flow, oh2012design, yu2010level, choi1991automatic, ramalingam2002sparse, hind1998assessing, chow1996effective, hardekopf2009semi}. 
This approach transforms top-level variables into Static Single Assignment (SSA) form, 
minimizing unnecessary points-to information propagations, a technique termed as semi-sparse flow-sensitive analysis. 
Later, the idea of stage analysis \cite{fink2008effective} was introduced. 
Those methods employ Andersen's analysis as a pre-analysis and convert both top-level variables and address-taken variables into the form of SSA to obtain a fully sparse pointer analysis \cite{hardekopf2011flow, li2011boosting}. 
The MTO-SS pointer analysis presented in this paper is built upon the fully sparse flow-sensitive analysis.

Field-sensitive analysis distinguishes different fields of objects to improve the accuracy of pointer analysis. 
The challenge of field-sensitive analysis for C/C++ lies in the fact that C/C++ allows taking the address of objects, 
and permits programs to access memory via store and load instructions using the obtained addresses. 
In order to solve this problem, Pearce et al. proposed PKH \cite{pearce2007efficient}, a method that assigns a unique identifier to each field of an object. 
This method extends Andersen's constraint graph with field identifiers as weights, generating new constraints to enhance the precision of pointer analysis. 
However, this method introduces a new type of cycles called positive weight cycles (PWCs), which are difficult to resolve. 
DEA \cite{lei2019fast} introduced a faster method to detect PWCs, aiming to enhance the performance of field-sensitive analysis. 
DSA \cite{lattner2007making} conducts accurate field-sensitive analysis when an object has only one type and degrades to field-insensitivity when the object has multiple types. 
Moreover, DSA is based on Steensgaard's unification-based \cite{das2000unification} pointer analysis, resulting in a rather coarse abstraction of objects. 
Min\'{e} \cite{mine2006field} demonstrates much higher accuracy with an approach that converts memory accesses to pointer arithmetic operations in an abstract interpretation framework, 
but this approach is unscalable to large programs.

cclyzer-ss \cite{balatsouras2016structure} combines Andersen's pointer analysis with the high-level structure information of objects to eliminate spurious points-to relationships. 
This method duplicates an object when a type conversion instruction is encountered and associates the type with this duplicated object. 
The duplicate object replaces the original object and continues to propagate. 
Furthermore, reverse propagation is also performed using the duplicate object to ensure the integrity of the points-to. 
However, this method leads to incorrect analysis results when analyzing objects that have multiple structures. 
This method also introduces significant performance overhead. 
The TYPECLONE technique is more refined, combining flow-sensitive analysis with the program's DWARF information. 
TYPECLONE \cite{barbar2020flow} leverages structural information from the program's debug information to enhance pointer precision. 
When an object is retrieved from memory by its type, this method duplicates the object and binds the type to this duplicate. 
If the same object is later accessed by a different type, another duplicate object is made with the new type bound to this new object. 
This method offers precise pointer analysis when an object retains a single type throughout its lifetime. 
However, when an object has multiple types throughout its lifetime, the method will produce inaccurate results. 
The performance overhead of this method is also notable. 
The method proposed in this paper for analyzing multi-structure objects not only handles the case where the object has more than one structure, but also improves the accuracy of pointer analysis. 

\section{Conclusion}

This paper presents a pointer analysis method for multi-structure objects called MTO-SS. 
The innovation of our method lies in introducing structure-flow-sensitivity to pointer analysis, which ensures the accuracy and completeness of the object's structure at different locations in the program. 
By incorporating the structural information of objects, our method makes the results of fully sparse flow-sensitive pointer analysis more accurate. 
Through experiments, we demonstrate that objects possessing multiple structural types are common. 
We evaluated the accuracy of MTO-SS. 
Its pointer analysis results surpass those of SPARSE and the TYPECLONE method. 





\bibliographystyle{ACM-Reference-Format}
\bibliography{sample-base}


\end{document}